\documentclass[twocolumn,showpacs,preprintnumbers,amsmath,amssymb]{revtex4}

\usepackage{graphicx}
\usepackage{dcolumn}
\usepackage{bm}

\begin{document}

\title{ Anomalous Properties of Quadrupole Collective States 
in $^{136}$Te and beyond }

\author{Noritaka Shimizu}
\email{shimizu@nt.phys.s.u-tokyo.ac.jp}
 \affiliation{Department of Physics, University of Tokyo, \\
Hongo, Bunkyo-ku, Tokyo, 113-0033, Japan\\
RIKEN, Hirosawa, Wako-shi, Saitama, 351-0198, Japan}

\author{Takaharu Otsuka}
\email{otsuka@phys.s.u-tokyo.ac.jp}
\affiliation{Department of Physics and Center for Nuclear Study, 
University of Tokyo, \\
Hongo, Bunkyo-ku, Tokyo, 113-0033, Japan\\
RIKEN, Hirosawa, Wako-shi, Saitama, 351-0198, Japan}

\author{Takahiro Mizusaki}
\email{mizusaki@nt.phys.s.u-tokyo.ac.jp}
\affiliation{
Institute of Natural Sciences,\\ Senshu University,
Higashimita, Tama, Kawasaki, Kanagawa, 214-8580, Japan}

\author{Michio Honma}
\email{m-honma@u-aizu.ac.jp}
\affiliation{
Center for Mathematical Science,\\ University of  Aizu,
Ikkimachi, Aizu-Wakamatsu, Fukushima, 965-8580, Japan}

\date{\today}

\begin{abstract}
The ground and low-lying states of neutron-rich exotic Te and Sn isotopes 
are studied in terms of the nuclear shell model by 
the same Hamiltonian used for the spherical-deformed shape phase 
transition of Ba isotopes, without any adjustment. 
An anomalously small value is obtained for $B(E2;0^+_1\rightarrow 2^+_1)$ 
in $^{136}$Te, consistently with a recent experiment.
The levels of $^{136}$Te up to yrast $12^+$ are shown to be in agreement 
with observed ones.   It is pointed out that $^{136}$Te can be an
exceptionally suitable case for studying mixed-symmetry 1$^+$, 2$^+$ and
3$^+$ states, and predictions are made for energies, M1 and E2 properties.
Systematic trends of structure of heavier and more exotic Sn and Te 
isotopes beyond $^{136}$Te are studied by Monte Carlo Shell Model, 
presenting an unusual and very slow evolution of collectivity/deformation.
\end{abstract}



\maketitle

\section{INTRODUCTION}

The nuclear collective motion is one of the central problems of 
nuclear structure physics.
In the nuclear shell model, 
a medium-heavy nucleus has many valence particles 
and these particles move collectively in a large single-particle space.
Because such a collective motion is dominated by quadrupole correlations, 
the corresponding states are referred to as the quadrupole collective states.
It is of great interest how such quadrupole collective states are
formed as one sails to more exotic regimes on the nuclear chart.
To explore this, a plausible approach is to adopt a Hamiltonian confirmed 
for its validity in and near stable regimes and apply it to unknown
regimes.  As such an attempt, in this paper, we shall discuss the structure of 
exotic Te isotopes with the neutron number ($N$) exceeding 82.
Because the proton number ($Z$) is 52 in Te isotopes 
and there are two valence protons with respect to the $Z$=50 closed core,
there should be certain proton-neutron correlations in such Te isotopes. 
A recent quantitative assessment of their structure, however, shows rather 
peculiar tendencies as will be presented. 

This paper is organized as follows.
We shall survey experimental situations and related empirical rules
in sect.~\ref{sec:exp}.  In sect.~\ref{sec:shell}, the shell-model 
Hamiltonian to be used is explained.  In sect.~\ref{sec:calc}, the 
calculation methods will be briefly over-viewed.
The structure of the exotic nucleus $^{136}$Te will be discussed in 
sect.~\ref{sec:136te}, with more specific discussions on 
mixed-symmetry states in sect.~\ref{sec:mixed} and 
on magnetic and quadrupole moments in sect.~\ref{sec:moments}.
Predicted systematic trends will be 
presented in sect.~\ref{sec:systematic}.  A summary will be given 
in sect.~\ref{sec:summary}.

\section{Experimental situation and empirical rules of quadrupole
collective states}
\label{sec:exp}

Certain basic properties of the quadrupole collective states can be
well described empirically by simple phenomenological
models.
For example, the systematic relation between the
excitation energy of the first $2^+$ state, $E_{2^+}$,
and the E2 transition strength from the ground $0^+$ state to
the first $2^+$ state, i.e.,
$B(E2)\uparrow$, has been studied well \cite{grodzins,raman}.
One of such useful formulas for this relation
is the modified Grodzins' rule \cite{raman},
which is written as:
\begin{equation}
  B(E2;0^+ \rightarrow 2^+) = (2.57 \pm 0.45)  E_{2^+}^{-1} Z^2 A^{-2/3} ,
  \label{eq:grodzins}
\end{equation}
where $E_{2^+}$[keV], $Z$, and $A$ denote the excitation energy of
$2^+_1$ state, the atomic number, and the mass number, respectively.
It has been confirmed \cite{raman} that a family of the Grodzins' rule
is extremely successful.

Another approach of the phenomenological relation can be found in
the systematic relation between  the above $B(E2)$ value
and $N_p \cdot N_n$, where $N_p$ and $N_n$ denote the numbers of
valence protons and neutrons, respectively \cite{hamamoto,casten1,casten2}.
Particularly, it has been stressed in \cite{casten1,casten2} that
this $B(E2)$ value can be given quite well as a function of the
quantity $N_pN_n$.
Since this $N_pN_n$ rule is quite robust in nuclei on and near the
$\beta$-stability line,
it is of great interest whether or not this $B(E2)$ value still follows
this rule, also in exotic nuclei far from the $\beta$-stability line.


While such empirical rules are successful to a good extent,
an exception has emerged in an experiment which extended the experimental
feasibility.  Namely, the anomalously small 
$B(E2;0^+_1\!\! \rightarrow \!\! 2^+_1)$ value 
of $^{136}$Te has been observed recently by Radford {\it et al.} \cite{teexp}. 
The $B(E2)$ value provided by the modified Grodzins' rule, 
eq.(\ref{eq:grodzins}), is $0.44(8)$ [e$^2$b$^2$], 
which is far from the experimental value,  0.103(15) [e$^2$b$^2$] \cite{teexp}.
The E2 transition rate is one of the most direct measures of 
the quadrupole deformation, and the fact that the rate for $^{136}$Te deviates
this much from the empirical rules is a challenge 
to the microscopic description of this nucleus.

\begin{figure}[htbp]
\begin{center}
\includegraphics[scale=0.3]{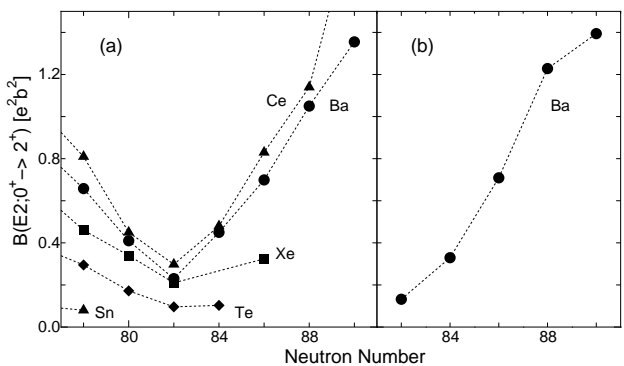}
\end{center}
\caption{ (a) Systematics of experimental 
   $B(E2;0^+_1\!\! \rightarrow \!\! 2^+_1)$ values 
   \cite{raman,teexp,xe136be2}, and 
   (b) calculated values for Ba isotopes \cite{baprl}.
  The triangle, circle, square, diamond and triangle symbols
  correspond to the $B(E2)$ values of Ce, Ba, Xe, Te and Sn isotopes, respectively.}
\label{be2all}
\end{figure}

Figure~\ref{be2all} shows observed 
$B(E2;0^+_1\!\! \rightarrow \!\! 2^+_1)$ values
of Sn, Te, Xe, Ba, and Ce isotopes.
Near the center of Fig.~\ref{be2all}(a) where $N=82$, 
the $B(E2)$ values are small, reflecting spherical ground states.
This value grows rapidly as one increases the number of neutron valence 
particles or holes.
Theoretical values for Ba isotopes \cite{baprl} are shown  
in Fig.~\ref{be2all}(b), demonstrating a rapid increase of the $B(E2)$
value in agreement with the experimental values.
The theoretical values were obtained by the Monte Carlo Shell Model
(MCSM) with the pair bases for a standard shell-model
Hamiltonian \cite{baprl}.
The calculated $B(E2)$ values of Ba isotopes are proportional to $N_n$ 
in the first approximation as suggested by Casten 
{\it et al.} \cite{casten2}.
However, the $B(E2)$ value for $^{136}$Te is only slightly larger
than the value for $^{134}$Te, in contrast to the trend 
of Ba isotopes.
As to theoretical approaches, Covello {\it et al.} made shell-model 
calculations based on a microscopic interaction as reported in \cite{teexp}, 
and Terasaki {\it et al.} have discussed this problem in terms of  
the quasiparticle random phase approximation (QRPA), while the pairing 
correlations are put in from the observed pairing gap \cite{teterasaki}.

\section{Shell model Hamiltonian}
\label{sec:shell}

We study the structure of nuclei around $^{136}$Te using the 
nuclear shell model.
The single-particle space and Hamiltonian for the shell model 
calculations are taken from existing ones which 
have been used successfully for a systematic description of
the shape phase transition in Ba isotopes from $N$=82 to 92,
which was already mentioned above \cite{baprl}.
The pairing correlation arises from the interplay between the 
single particle energies and the pairing interaction.  

This shell model Hamiltonian is different from 
the one used by Covello {\it et al.} \cite{teexp}.   
In fact, Covello  {\it et al.} derived a realistic effective interaction 
from the bare Nucleon-Nucleon ($NN$) interaction.

The present single-particle model space consists of the valence orbits 
in the $Z=50\sim 82$ proton shell and those in the $N=82\sim 126$ 
neutron shell.

\begin{figure}[htbp]
\begin{center}
\includegraphics[scale=0.4]{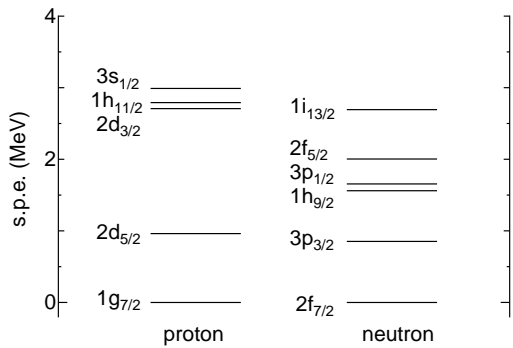}
\end{center}
\caption{ Proton (left) and neutron (right) single-particle orbits and
their energies.  The energies are taken from experiments 
 \cite{speSb,speSn}.}
\label{npspe}
\end{figure}

The Hamiltonian we shall use is comprised of the three parts, 
\begin{equation}
  H =H_\pi + H_\nu + V_{\pi\nu}
\end{equation}
where $H_\pi$ ($H_\nu$) means the proton (neutron) Hamiltonian 
and $V_{\pi\nu}$ denotes a proton-neutron interaction.
The $H_\pi$ ($H_\nu$) includes proton (neutron) 
single-particle energies and a two-body 
interaction between valence protons (neutrons). 
The proton (neutron) single-particle energies are taken from 
experimental levels of $^{133}$Sb ($^{133}$Sn) \cite{speSb} (\cite{speSn}).
These single-particle orbits and their energies are shown in 
Fig.~\ref{npspe}

The two-body interaction includes the monopole and quadrupole 
pairing interactions and the quadrupole-quadrupole interaction.
The values for protons (neutrons) are 
$g^{(0)} = 0.21(0.13)$ MeV, 
$g^{(2)} = 0.22(0.14)$ MeV,  and $f^{(2)} = -0.0002(0.0002)$ MeV/fm$^4$,
where $g^{(0)}$, $g^{(2)}$, and $f^{(2)}$ are strength parameters of 
the monopole and quadrupole pairing interactions and the 
quadrupole-quadrupole interaction, respectively \cite{baprl}.
The interaction between a proton and a neutron
is assumed to be of quadrupole-quadrupole type
with its strength $f^{(2)}_{\pi\nu} = - 0.0014$ MeV/fm$^4$ \cite{baprl}.
Although the present shell-model Hamiltonian is schematic to a certain
extent, it has been tested as being successful in reproducing quadrupole
collective states of Ba isotopes over the shape phase transition.
It is of a great interest to see whether such a Hamiltonian can
be still valid for the study of the anomalously small $B(E2)$ value
of $^{136}$Te. 

We use the same effective charges as in the calculation for Ba
isotopes: effective charges are 
$e_p = 1.6e$ and $e_n = 0.6e$ for proton and neutron, respectively.
We calculate magnetic transitions with standard $g$-factors as
we shall show later.

\section{Conventional and Monte Carlo shell model calculations}
\label{sec:calc}

The structure of the nucleus $^{136}$Te is studied by the conventional
shell-model diagonalization for the Hamiltonian discussed in the
previous section.  The OXBASH code is used \cite{oxbash}.

For heavier Te isotopes, however, a larger dimension of the Hilbert space 
prevents us more and more severely from diagonalizing its Hamiltonian matrix.
In order to overcome such a growing difficulty, 
the Monte Carlo Shell Model (MCSM) has been proposed \cite{qmcd1,qmcd4,ppnp}, 
which enabled us to apply the large-scale shell model calculation 
also to the collective states of the medium-heavy nuclei.
For the study of quadrupole collective states
in even-even nuclei, the most crucial dynamics is the competition
between the quadrupole deformation and the pairing correlation
\cite{ringschuck}.
In order to handle such situations, the MCSM with pair bases has
been introduced and has been successfully applied to the description
of the shape phase transition in Ba isotopes with $N > 82$ \cite{baprl}.
In addition, even to the case of $^{136}$Te, MCSM has been used for the 
analysis of pair structure, because the OXBASH code does not have such 
a capability.

We note that a preliminary and very brief report of a part of the following
results has been presented in \cite{wyoming}.



\section{Levels of $^{136}$Te}
\label{sec:136te}

\begin{figure}[tbp]
\begin{center}
\includegraphics[scale=0.4]{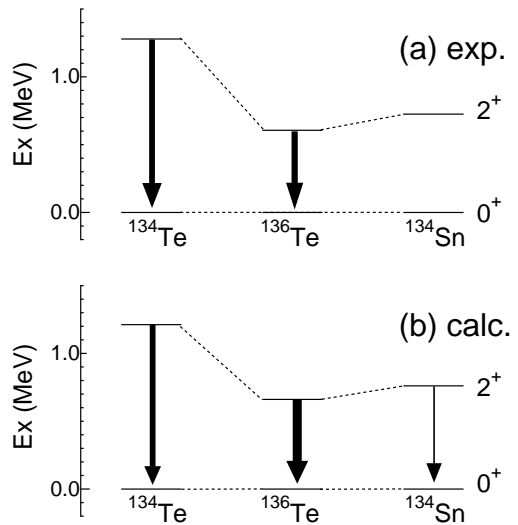}
\end{center}
\caption{ Excitation energy of $2^+_1$ and B(E2) of $^{134,136}$Te and $^{134}$Sn. 
The upper part is obtained experimentally \cite{teexp}, 
while the lower part is calculated by 
the present work. 
The arrow widths are proportional to the $B(E2)$ values.}
\label{fig:sntelevels}
\end{figure}

We first discuss how the $0^+_1$ and $2^+_1$ wave functions of $^{136}$Te
are constructed. 
Figure \ref{fig:sntelevels} shows
the $2^+_1$ level of $^{136}$Te, together with those of  
$^{134}$Te and $^{134}$Sn. 
The nucleus $^{136}$Te has two valence protons and two neutrons, while  
the neighboring nuclei, $^{134}$Te and $^{134}$Sn, 
have two valence protons or two valence neutrons, respectively.
We analyze wave functions of the $0^+_1$ and $2^+_1$ states in
terms of shell model with these valence nucleons.

The ground state wave function of $^{134}$Sn is 
written as 
\begin{equation}
  |S_{\nu} \rangle = S^\dagger_{\nu} |- \rangle,
\end{equation}
where 
$|-\rangle$ indicates the inert core ({\it i.e.} $^{132}$Sn) and 
$S^\dagger_{\nu}$ denotes the creation operator of a pair of valence neutrons  
coupled to the angular momentum 0.  The $S^\dagger_{\nu}$ operator is
defined as  
\begin{equation}
  S^\dagger_{\nu} \equiv \sum_{j} \alpha_j \left[c^\dagger_j \times 
  c^\dagger_j \right]^{(0)} ,
\end{equation}
where $c^\dagger_j$ denotes the creation operator of a neutron in a 
single-particle orbit $j$, and 
$\alpha_j$ indicates an amplitude giving the proper normalization of the
state $|S_{\nu} \rangle$.
The values of  $\alpha_j$'s are 
determined by the diagonalization of the Hamiltonian matrix.
The ground state wave function of $^{134}$Te is 
written similarly as 
\begin{equation}
  |S_{\pi} \rangle = S^\dagger_{\pi} |- \rangle,
\end{equation}
with $S^\dagger_{\pi}$ defined correspondingly. 

Likewise, the $2^+_1$ state of $^{134}$Sn 
is provided by a $2^+$ state of two neutrons, called $D_\nu$ pair,
on top of the $^{132}$Sn core.  Similarly, 
the $2^+_1$ state of $^{134}$Te is given by the $D_\pi$ pair.  
These D pairs are created by the operators,  
\begin{equation}
  D^\dagger_M \equiv \sum_{jj'} \beta_{jj'} \left[c^\dagger_j 
    \times c^\dagger_{j'} \right]^{(2)}_M,
\end{equation}
where the subscript $\pi$ or $\nu$ is omitted for brevity,  
$M$ means the $z$-component of angular momentum, and 
$\beta_{jj'}$ stands for amplitude.
The values of $\beta_{jj'}$ are determined by the diagonalization of 
the Hamiltonian matrix for the state 
$|D \rangle _M \equiv D^\dagger_M |-\rangle $, 
so that it is properly normalized.
We shall omit $M$ hereafter because it is not essential.
These S- and D-pairs are usually called collective pairs, because they 
are comprised of coherent superposition of various nucleon pairs,
although the coherence can be modest in the following cases. 

Figure~\ref{fig:sntelevels} shows that the first 2$^+$ level is
quite well reproduced by the present Hamiltonian. 
The $B(E2;0^+_1\!\! \rightarrow \!\! 2^+_1)$ value 
is $0.096$ [e$^2$b$^2$] and $0.027$ [e$^2$b$^2$] for $^{134}$Te and 
$^{134}$Sn, respectively.  Experimentally, only the former is known as
$0.096 (12)$ [e$^2$b$^2$] \cite{teexp}, in a reasonable agreement with the 
present calculation and also with the results in 
\cite{teexp,teterasaki}.
For $^{134}$Sn, the $B(E2;0^+_1\!\! \rightarrow \!\! 2^+_1)$ value 
becomes $0.035$ [e$^2$b$^2$] in the shell-model calculation by 
Coraggio {\it et al.} \cite{134SnCoraggio}, whereas the QRPA result by
Terasaki {\it et al.} \cite{teterasaki} gives a considerably smaller value.
The present value is in between and closer to the former one. 
The Nilsson result in \cite{teterasaki} seems to resemble the two 
shell-model values. 

The shell-model wave functions of the $0^+_1$, $2^+_1$ and $2^+_2$ states 
of $^{136}$Te can be written as,
\begin{eqnarray}
  \label{eq:wftezero}
  |0^+_1 \rangle &=& 0.91 \times  |S_\nu \times S_\pi \rangle + 
                            \cdot\cdot\cdot , \\
  \label{eq:wftefirsttwo}
  |2^+_1 \rangle &=& 0.82 \times  |D_\nu \times S_\pi \rangle 
  + 0.45 \times  |S_\nu \times D_\pi \rangle + \cdot\cdot\cdot ,\\
  \label{eq:wftesecondtwo}
  |2^+_2 \rangle &=& 0.38 \times  |D_\nu \times S_\pi \rangle 
  - 0.76 \times  |S_\nu \times D_\pi \rangle + \cdot\cdot\cdot , 
\end{eqnarray}
where ``$\cdot \cdot \cdot $'' means other minor components and 
$|S_\nu \times S_\pi \rangle \equiv S^\dagger_\nu S^\dagger_\pi 
  |- \rangle $, etc. 
Equation (\ref{eq:wftezero}) implies that the $0^+_1$ state  
is accounted for by the state $|S_\nu \times S_\pi \rangle $ up to
83 \% in probability.

Moving to the first $2^+$ state, eq.(\ref{eq:wftefirsttwo}) indicates 
that the probability of the component $|D_\nu \times S_\pi \rangle$ 
is larger by a factor of about four than that of 
$|S_\nu \times D_\pi \rangle $.  
This asymmetry is rather unusual for the
first $2^+$ state of nuclei with open shells for protons and neutrons; 
strong proton-neutron couplings mix protons and neutrons more equally 
in other usual (maybe stable) nuclei,  
giving rise to a more symmetric superposition.
Figure~\ref{fig:sntelevels} (b) shows that the excitation energy of the 
$D_\nu$ state measured from $S_\nu$ is 0.76 MeV, which is about 0.45 keV 
lower than the excitation energy  (1.21 MeV) of the $D_\pi$ relative 
to the $S_\pi$ state.
The origin of the above asymmetry in eq. (\ref{eq:wftefirsttwo}) 
is nothing but this difference in the excitation energies 
of $D_\pi$ and $D_\nu$.
If the proton-neutron correlation is strong enough,
such a difference is overcome, and protons and neutrons move in 
coherent manners as is the case, for instance, with heavier Ba isotopes 
with the same Hamiltonian \cite{baprl}.  However, because of fewer valence
nucleons, this is not the case in $^{136}$Te, and  
the difference between proton and neutron remains crucial, 
yielding the asymmetry in the wave function in eq. (\ref{eq:wftefirsttwo}).  
The small excitation energy of the $D_\nu$ state is clearly due to
the weaker monopole pairing between neutrons ($g^{(0)} = 0.13$ MeV) than 
the pairing between protons ($g^{(0)} = 0.21$ MeV).  
Although the quadrupole pairing interaction follows the same trend
and the difference in the monopole pairing is partly canceled by 
the quadrupole pairing, 
the $D_\nu$ state is still lower than the $D_\pi$ state.  

This asymmetry in eq.(\ref{eq:wftefirsttwo})
decreases the proton-neutron coherence in the E2 transition 
from the ground to the $2^+_1$ state, 
resulting in a weaker E2 transition.
In addition, the dominant weight of the $|D_\nu \times S_\pi \rangle$ 
state makes the $B(E2)$ value further smaller because of the small 
effective charge (0.6$e$) for neutrons.  
Thus, $B(E2;0^+_1\!\! \rightarrow \!\! 2^+_1)$ value 
becomes $0.15$ [e$^2$b$^2$] for $^{136}$Te.
This value is larger than the $^{134}$Te value only by a rather 
modest factor, about 1.5, consistently with the experimental
observation.  In fact, 
this value appears to be slightly larger than the experimental value,  
$0.103 (15)$ [e$^2$b$^2$], 
reported by Radford {\it et al.} \cite{teexp}. 
On the other hand, the present value is smaller than the theoretical
value, $0.25$ [e$^2$b$^2$], by Covello {\it et al.} \cite{teexp}, 
although their calculation was made based on a fully microscopic
$NN$ interaction \cite{covello}.
We note that a smaller value, 0.16 [e$^2$b$^2$], has been reported 
later in \cite{teexp_aizu}
by the same authors as those of \cite{teexp} as a result of a more
consistent calculation still within the same microscopic interaction.
The present value is closer to the value by Terasaki 
{\it et al.} obtained by a QRPA calculation using 
observed pairing gaps \cite{teterasaki}.

\begin{figure}[tbp]
\begin{center}
\includegraphics[scale=0.45]{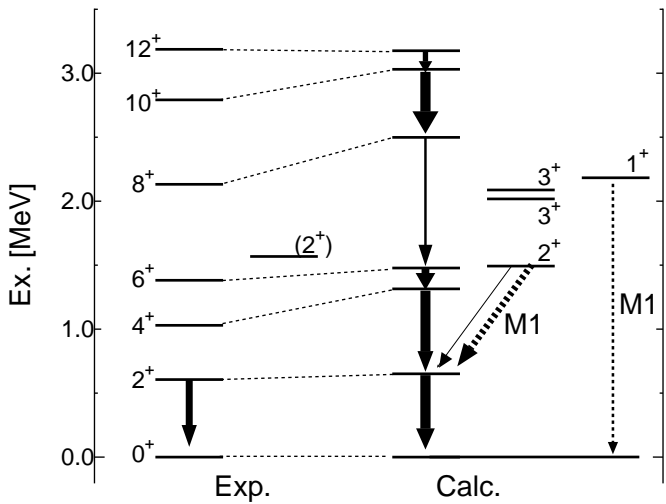}
\end{center}
\caption{Level schemes of $^{136}$Te obtained by the experiments 
\cite{teexp,tedata,tedata1,tedata2} 
and the present shell-model calculation.
The solid arrows indicate E2 transitions with widths of proportional to 
the $B(E2)$ values (note that calculated $B(E2;2^+_1\rightarrow0^+_1)$
=$0.030$ [e$^2$b$^2$]).  
The dashed arrows represent M1 transitions with widths of proportional to 
the $B(M1)$ values (see the text). 
 }
\label{fig:te136level}
\end{figure}

We shall move on to higher states, as one of the advantages  
of the shell model calculation is the capability of studying higher 
and/or side states.
Figure~\ref{fig:te136level} shows a level scheme of $^{136}$Te
as compared to experiment \cite{teexp,tedata,tedata1,tedata2}. 
The even-spin yrast levels are shown up to 12$^+$.  The excitation 
energy is well reproduced, while the levels somewhat deviate  
for the 4$^+$, 8$^+$ and 10$^+$ states. The calculated 4$^+$ level is
higher, mainly because the Hamiltonian was designed not to include 
the hexadecupole pairing, for simplicity.  
The 6$^+$ state is comprised mainly of the 6$^+$ pair of neutrons in 
$2f_{7/2}$ and the $S_{\pi}$ pair.  Since this state has nothing to
do with the hexadecupole pairing, it 
exhibits a good agreement to experiment.
The 8$^+$ and 10$^+$ states, on the other hand, should contain 4$^+$ pairs 
in their wave functions resulting in certain deviations.
The difference of wave function contents between the 6$^+$ and 8$^+$ states
should be the origin of the almost vanishing $E2$ transition between them.

\section{Mixed-symmetry states in $^{136}$Te}
\label{sec:mixed}

The structure of the $2^+_2$ state is quite interesting.
Equation (\ref{eq:wftesecondtwo}) shows that  
this state contains considerable amount of the $|S_\nu \times D_\pi \rangle $ 
state as well as $|D_\nu \times S_\pi \rangle$     
with the opposite signs.
The state $
\{ |D_\nu \times S_\pi \rangle - |S_\nu \times D_\pi \rangle \} / \sqrt{2} $
is clearly anti-symmetric with respect to interchanges between proton 
pairs and neutron pairs, and is called a mixed-symmetry state
\cite{ibm2,PhD,otsuka_book}. 
Although the mixed-symmetry states are defined with the
IBM-2, the bosons and the collective pairs can be mapped onto each other 
\cite{ibm2,PhD,otsuka_book}, and the concept of
the mixed-symmetry states will be used in this context.
The $2^+_2$ state ($|2^+_2 \rangle$ in eq. (\ref{eq:wftesecondtwo})) 
is dominated by this mixed-symmetry state up to 65 \%.  
Its excitation energy is about 1.5 MeV, 
as shown in Fig.~\ref{fig:te136level}.  The mixed-symmetry states lie
usually in the energy region of high level density, and therefore it is
difficult to identify them.  In the present case, the situation may be
more favorable for its identification.
The $2^+_2 \rightarrow 2^+_1$ M1 transition is rather strong with
$B(M1) $ = 1.04 $[\mu_n^2]$, and dominates the $2^+_2 \rightarrow 2^+_1$
transition because $B(E2; 2^+_2 \rightarrow 2^+_1)$ is as small as
0.001 [e$^2$b$^2$].  This is a consequence of the fact that the 
$2^+_1$ and $2^+_2$ states have opposite proton-neutron phase contents  
(See eqs. (\ref{eq:wftefirsttwo}) and (\ref{eq:wftesecondtwo})) and the
M1 transition has a strong isovector part.  
While there are several tentative 2$^+$ states
in experiment in the energy region of the calculated $2^+_2$ state,  
the lowest one is indicated in Fig.~\ref{fig:te136level}. 
The calculated $B(E2; 0^+_1 \rightarrow 2^+_2)$ is
0.03 $[e^2 b^2]$, which is one fifth of the $B(E2; 0^+_1 \rightarrow 2^+_1)$, 
due to the cancellation between proton and neutron contributions.

The calculated $1^+_1$ and $3^+_{1,2}$ states are shown also 
in Fig.~\ref{fig:te136level}.
The relevant mixed-symmetry states are of the type 
$|D_\nu \times D_\pi \rangle$.  Namely, if $|D_\nu \rangle$ and 
$|D_\pi \rangle$ are coupled to an odd angular momentum, the wave 
function becomes anti-symmetric with respect to the interchange between 
$|D_\nu \rangle$ and $|D_\pi \rangle$ and can be called of mixed-symmetry
\cite{ibm2}. 
The $1^+_1$ state has the overlap probability of 76 \% with the
$|D_\nu \times D_\pi ; J=1 \rangle $ with $J$ being the total angular
momentum.  The corresponding probability is fragmented 
as 23 \% and 51 \% for the $3^+_1$ and
$3^+_2$ states, respectively, and both of them are shown 
in Fig.~\ref{fig:te136level}.
The excitation energies of $1^+$ and $3^+$ $|D_\nu \times D_\pi \rangle$ 
states are expected to be about equal to the sum of 
the excitation energies of the 
$2^+_1$ and $2^+_2$ states, as is true for the IBM-2 cases without so-called
Majorana interaction \cite{PhD}.
This feature is maintained in Fig.~\ref{fig:te136level} 
despite mixed impurities in actual eigenstates.

The calculated $B(M1; 0^+_1 \rightarrow 1^+_1) $ turns out to be 
1.14 $[\mu_n^2]$, which is
rather strong as a measure of mixed symmetry states, although this M1 
transition contains spin transition as well as orbital one.
There are many experimental levels in the same energy region, but 
they are not shown in Fig.~\ref{fig:te136level} because their spin/parity 
assignment is presently unavailable.

Thus, the present shell model calculation exhibits the full set of the 
mixed-symmetry states, 
$1^+$, $2^+$ and $3^+$, in low-excitation energy region. 
The experimental identification of the full members of these 
$1^+$, $2^+$ and $3^+$ mixed-symmetry states has been proposed only 
for a few nuclei, for instance, $^{94}$Mo \cite{94Mo}.
The mixed-symmetry states are pushed too
high in the cases with strong proton-neutron correlations which 
certainly favor coherent couplings of protons and neutrons.  In exotic
nuclei like $^{136}$Te, this may not be the case.  
Thus, with $^{136}$Te, one may be able to identify the mixed-symmetry 
states and investigate their various aspects.

\section{Moments of $^{136}$Te}
\label{sec:moments}

We next discuss properties of magnetic and quadrupole moments of
$^{136}$Te.

\begin{figure}[tbp]
\begin{center}
\includegraphics[scale=0.5]{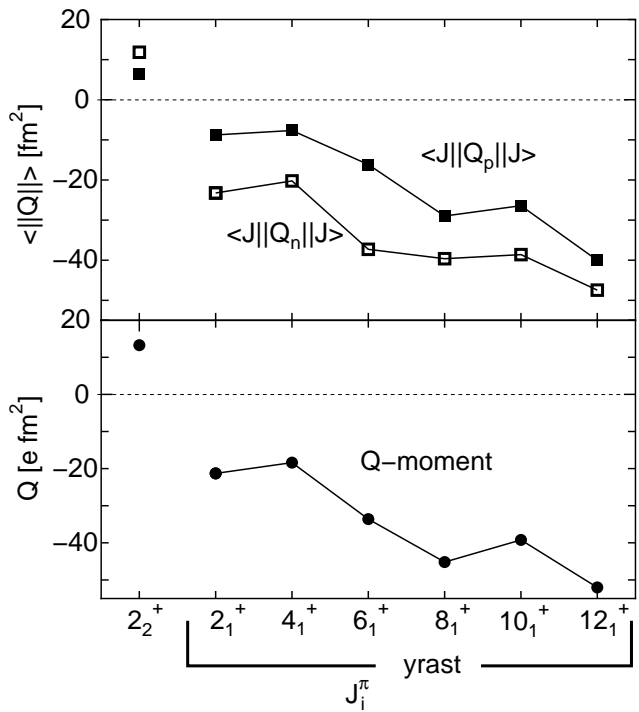}
\end{center}
\caption{
  Calculated reduced quadrupole matrix elements of protons 
  ($\langle J^+||Q_\nu||J^+\rangle$) and 
  neutrons ($\langle J^+||Q_\pi||J^+\rangle$) [${\rm fm}^2$] 
  and spectroscopic electric quadrupole moments [$e \, {\rm fm}^2$] 
  for $^{136}$Te.
  The states are, from left to right,   
  $2^+_2$, $2^+_1$, $4^+_1$, $6^+_1$, $8^+_1$, $10^+_1$, and $12^+_1$. 
  }
\label{qmom_te136}
\end{figure}

Figure \ref{qmom_te136} shows 
reduced matrix elements ($\langle J ||Q_\nu|| J \rangle $, 
and $\langle J ||Q_\pi|| J \rangle $) of quadrupole operators 
as well as the spectroscopic quadrupole moments.  
The $2^+_1$ and $4^+_1$ states show small values.
In the yrast states, all the matrix elements of neutrons are 
larger in magnitude than the corresponding ones of protons, 
because the yrast states are dominated by the neutron excitations.
The same quantities of the $2^+_2$ state are shown at the left end of 
Fig.\ref{qmom_te136}, exhibiting a weak oblate deformation.

\begin{figure}[tbp]
\begin{center}
\includegraphics[scale=0.4]{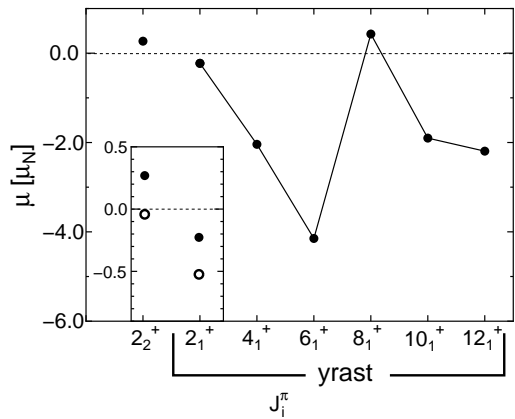}
\end{center}
\caption{Magnetic dipole moments of low-lying excited states 
  ($2^+_2$, $2^+_1$, $4^+_1$, $6^+_1$, $8^+_1$, $10^+_1$, and $12^+_1$) 
   of $^{136}$Te.
  The open circles in the inset mean the result with the g-factors
  used by Honma {\it et al.} \cite{gxpf1}.}
\label{mag_te136}
\end{figure}

Figure \ref{mag_te136} shows the magnetic dipole moments. 
The orbital and spin g-factors are taken as
$(g_{l\nu},g_{l\pi}) = (0.0, \, 1.0)$ 
and $(g_{s\nu},g_{s\pi}) = (-2.674, \, 3.906)$.
The spin ones are quenched by a factor 0.7 
from the free spin $g$-factors, $(g_{s\nu},g_{s\pi}) = (-3.82, 5.58)$.
We now discuss the magnetic dipole moment of the $2^+_1$ state of 
$^{136}$Te.  The wave function in eq. (\ref{eq:wftefirsttwo})
suggests that the two valence neutrons in this state are coupled 
primarily to the angular momentum two, while the two valence protons
are coupled mostly to zero.  The magnetic moment of the $2^+_1$ state, 
therefore, comes mainly from neutrons.
On the other hand, the orbital and spin g-factors of the neutron are 
zero and negative, respectively.  Combining all these facts,
it is deduced that the magnetic dipole moment of the $2^+_1$ state 
is most likely negative.  Figure \ref{mag_te136} confirms that this
is the case.   
In contrast, the magnetic moment takes a small positive value 
for the $2^+_2$ state, owing to the orthogonal structure. 
This trend does not change basically by 
using other reasonable sets of g-factors.  For instance,
the spin quenching 0.9 and the orbital isovector correction 0.1 were
used for $pf$-shell nuclei by Honma {\it et al.} \cite{gxpf1}. 
Figure \ref{mag_te136} (inset) indicates a negative overall shift with 
this set.
The present result for the moment of the $2^+_1$ state 
resembles the QRPA result ($-0.174$) \cite{teterasaki}.
The magnetic moments of the yrast states exhibit monotonic increase 
up to the $6^+_1$ state, and a different structure sets in
as expected from the level scheme in Fig.~\ref{fig:te136level}.

\section{Systematic trends in heavier Te and Sn isotopes}
\label{sec:systematic}

We shall now look at systematic trends predicted by the same Hamiltonian
as we explore into more exotic regions of heavier Sn and Te isotopes.

\begin{figure}[tbp]
\begin{center}
\includegraphics[scale=0.4]{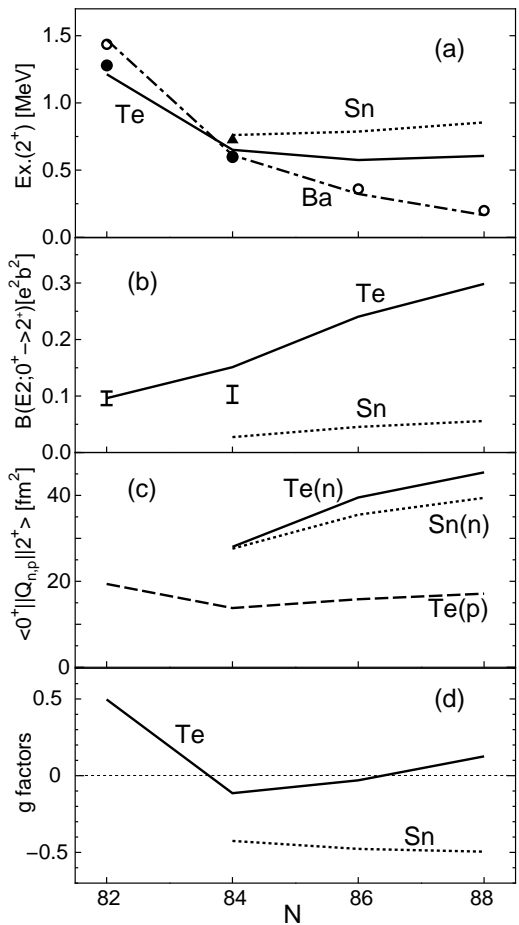}
\end{center}
\caption{ 
  Properties of Sn and Te isotopes as a function of the neutron number, $N$.
  (a) Excitation energies of $2^+$ states.
      The triangle and filled circles denote the experimental values 
      for Sn and Te, respectively \cite{teexp}, while the dotted and
      solid lines are calculated values for Sn and Te, respectively.
      The $2^+$ levels of Ba isotopes are shown by open circles 
      (experiment) and by dashed-dotted line (calculation).
  (b) $B(E2;0^+_2 \rightarrow 2^+_1)$ values.
      The bars are experimental data \cite{teexp}, while lines are
      calculations.
  (c) Reduced matrix elements of quadrupole operators.
  (d) Calculated g-factors of the 2$^+_1$ state.
}
\label{fig:teall}
\end{figure}

In Fig.~\ref{fig:teall}(a), the calculated excitation energies of 
$2^+_1$ states of Sn and Te isotopes are plotted as a function of $N$.
The $2^+_1$ level of Sn isotopes stays almost constant, while it
goes up slightly for larger $N$. 
This constancy is a common feature of semi-magic nuclei, but
should be examined experimentally.
On the other hand, the $2^+_1$ level of Te isotopes comes down
at the beginning, but again stays constant after $N$=84.
This is rather unusual, because the $2^+_1$ level continues to
go down in most of medium-heavy open-shell even-even nuclei.
This nearly constant level systematics contradict  
the empirical predictions \cite{casten1,casten2} also.
Such unusual trend may become more prominent in (some) further exotic 
nuclei where proton-neutron coupling is even weaker.  

Certainly, by increasing
the number of valence protons, the same proton-neutron interaction can 
promote stronger deformation, static or dynamic, and ``canonical''
collective motions should set in.  An example of this, 
Fig.~\ref{fig:teall}(a) includes the $2^+_1$ levels of Ba
isotopes calculated by the same Hamiltonian \cite{baprl}.  
These calculated levels are very close to the experimental ones.
The $2^+_1$ level of Ba isotopes indeed keeps falling down 
as $N$ increases.  

Figure \ref{fig:teall}(b) shows the $B(E2;0^+_1 \rightarrow 2^+_1)$ values
of Sn and Te isotopes.  The value for $^{136}$Te has been discussed in 
sect.~\ref{sec:136te}.
This $B(E2)$ value of Sn isotopes increases very slowly.  This 
behavior is similar to lighter Sn isotopes with $N < 82$ as a function
of the number of neutron holes.
The $B(E2)$ value can be expected to increase linearly as a function 
of the valence neutron number, $N_n$ (=$N-82$ in this case), 
in a picture of the simple boson model \cite{ibm2,PhD,otsuka_book}, while 
this $B(E2)$ is somewhat suppressed due to the Pauli blocking
\cite{Mizusaki}.  
This ``spherical $N_n$ effect'' will be discussed once again.

In contrast, the $B(E2)$ value of Te isotopes increases relatively faster.
The difference from the value of $^{134}$Te fits well to a linear increase
as a function of $N_n$.  Namely, 
the theoretical prediction is somewhat consistent with the model of 
Casten {\it et al.} \cite{casten1,casten2}.
Experimental investigations are of great interest.

Figure \ref{fig:teall}(c) shows the reduced matrix elements of 
quadrupole operator between the $0^+_1$ and $2^+_1$ states 
for Te and Sn isotopes.  No effective charges are included.
For Te isotopes, the contributions of protons and neutrons 
are separated, whereas neutrons are the only valence particles in Sn.  
The tendency of the neutron matrix elements of Te isotopes
is similar to those of Sn isotopes, while the presence of valence protons 
enlarges the neutron matrix elements of Te isotopes to a certain extent.

We point out that the proton matrix element in Fig.~\ref{fig:teall} (c) 
decreases from $^{134}$Te to $^{136}$Te.  This happens because 
the $2^+_1$ wave function is dominated by $|D_\nu \times S_\pi 
\rangle$, whereas only $|S_\nu \times D_\pi \rangle$ can be excited by
the proton quadrupole transition from $|S_\nu \times S_\pi \rangle$.  
In this picture, the proton matrix element can be about a 
half of that of $^{136}$Te, because of
the small amplitude of $|S_\nu \times D_\pi \rangle$ component in
eq.~(\ref{eq:wftefirsttwo}).  The decrease is, however, only by about 20 \%, 
owing to re-arrangements of other minor components of the $0^+_1$ and 
$2^+_1$ wave functions so as to enhance quadrupole collectivity.

The proton contribution increases only modestly as a function of 
$N_n$ in Fig.~\ref{fig:teall} (c).  The $B(E2)$ value of Te isotopes 
increases mainly due to the increase of the neutron matrix element as 
the ``spherical $N_n$ effect'' mentioned above.
Thus, the evolution of the collectivity/deformation in Te isotopes is
mainly due to neutron part of the wave function.  The proton part seems
to be saturated already at $^{136}$Te.  The evolution driven only by 
neutrons seems to be rather slow.  In fact, one can compare the growth
of the $B(E2)$ value of Te isotopes to that of Ba isotopes shown 
in Fig.~\ref{be2all}.  The $B(E2)$ of Ba isotopes grows so rapidly that
it overscales Fig.~\ref{fig:teall} (b).  This difference is due to the fact 
that both proton and neutron wave functions undergo the phase transition 
from the spherical to deformed intrinsic structures in Ba isotopes, 
and {\it both} proton and neutron matrix elements become larger
as the neutron number approaches 90.  On the other side, both proton and 
neutron wave functions remain basically spherical in Te isotopes and 
the evolution reflects only the ``spherical $N_n$ effect''.

Figure \ref{fig:teall}(d) shows g-factors of the $2^+_1$ state of Sn and Te 
isotopes.  The g-factor of Te isotopes shows a weak tendency to the
collective value, $Z/A$ \cite{BM}, or IBM-2 value, $N_p / (N_p + N_n)$
\cite{IBM-g}.

\section{Summary}
\label{sec:summary}

The structure of an exotic nucleus $^{136}$Te and its vicinity
has been studied by the shell model, using the MCSM technique.
The unusually small value of 
$^{136}$Te $B(E2; 0^+_1 \rightarrow 2^+_1)$ has been explained
without any adjustment. 
Based on weak proton-neutron coupling in $^{136}$Te, mixed-symmetry
properties are discussed, proposing this nucleus as an excellent play
ground for this subject. 
We also provide with predictions of Te isotopes beyond $^{136}$Te.
The evolution of the collective motion as a function of the neutron
number may be rather different from that in more stable nuclei, and 
a slow growth of the collectivity is predicted, which deviates from
empirical predictions.
The calculations for heavier Te isotopes are already huge, and 
have been carried out by the MCSM.

\section{acknowledgments}

The authors acknowledge Professor A.~Gelberg for reading the manuscript.
This work was supported mainly by Grant-in-Aid for Specially
Promoted Research (13002001) from the MEXT, and by
the RIKEN-CNS collaboration project
on large-scale nuclear structure calculation.
N.S. acknowledges the Special Postdoctoral Researchers Program of RIKEN 
(Grant No. B55-52050).
The conventional shell-model calculation was carried out by
the code {\sc oxbash} \cite{oxbash}.

\end{document}